# Fast learning synapses with molecular spin valves via selective magnetic potentiation

Alberto Riminucci[1], Robert Legenstein[2]

[1] *Institute for the study of Nanostructured Materials, CNR, Bologna, Italy, alberto.riminucci@cnr.it*

[2] *Institute for Theoretical Computer Science, Graz University of Technology, Graz, Austria, legi@igi.tugraz.at*

**Abstract**

We studied LSMO/Alq3/AlOx/Co molecular spin valves in view of their use as synapses in neuromorphic computing. In neuromorphic computing, the learning ability is embodied in specific changes of the synaptic weight. In this perspective, the relevant parameter is the conductance of the molecular spin valve, which plays the role of the synaptic weight. In this work we demonstrated that the conductance can be changes by the repeated application of voltage pulses. We studied the parameter space of the pulses in order to determine the most effective voltage and duration of the pulses. The conductance could also be modified by aligning the magnetizations of the ferromagnetic electrodes parallel or anti parallel to each other. This phenomenon, known as magnetoresistance, affects high conductance devices while leaving low conductance devices unaffected. We studied how this weight update rule affected the speed of reward-based learning in an actor-critic framework, compared to a linear update rule. This nonlinear update performed significantly better (50 learning trials; Epochs to reach a performance goal of 0.975 was 896±301 in the nonlinear case and 1076±484 in the nonlinear case; Welch's t-test: $p<0.05$). The linear update resulted in more learning trails with very long convergence times, which was largely absent in the nonlinear update.

There is tremendous potential in technologies that can endow small, portable objects with a degree of intelligence. Also known as edge computing[1], it is the next step to follow after the introduction of the Internet of Things, whereby computation is carried out close to the sensor node rather than in the cloud. Devices based on neuromorphic edge computing could adapt to a changing environment and to changing demands. They could do so in a coordinated manner, so as to increase no only efficiency locally, but also globally.

For these reasons it is vital to develop simple technologies that can implement artificial intelligence with minimal energy consumption[2]. One of the most promising devices in this sense is the memristor[3]. This device is ideally suited to act at the synaptic weight in a neuromorphic chip. Its non-volatility means that no energy must be spent in keeping its stored value. If arranged in a crossbar, its node size is $4f^2$ so that very high device densities can be achieved on a chip[4].

The effort devoted to the development of the memristor is considerable. Many active materials at this time are based on binary oxides, in which mobile oxygen vacancies give them the ability to change the conductance[5]. In synaptic terms, a high conductance corresponds to a high weight synapse while a low conductance one corresponds to a low weight one. In most circuit architectures, such as the cross bar, the memristors must be addressed individually. Learning cannot be parallelized, and this can be a source of inefficiency. If, for example, some synapses are to be pruned (i.e. depressed) or potentiated, it must be done separately for each one.

Molecular spin valves have a unique advantage over memristors. While their conductance can be controlled electrically, as with memristors, it can also be controlled magnetically[6]. The magnetic control can be applied simultaneously to all synapses and, crucially, potentiated synapses respond to it, while depressed ones do not. This means that the magnetic field can selectively address only potentiated synapses, for example by making them more potentiated still. There are very few instances of the concomitant presence of intertwined GMR and resistive multistability[7]–[15]. On the one hand, the highest GMRs are achieved by using MgO tunnel barriers[16]. On the other hand, the presence of some kind of defect or impurity is required for the existence of resistive multistablity. Finally, their interplay is far from obvious. Several systems can give resistive multistablity and GMR at low temperature, but examples of devices working at ambient temperature are scarce[9], [14], [17], [18].

In this article we demonstrate that La0.7Sr0.3MnO3(LSMO)/ Tris-(8-hydroxyquinoline)aluminium(Alq3)/ AlOx/Co behave as a synapse when subjected to a voltage pulse. We also demonstrate that the ability of the magnetic field to selectively address potentiated (i.e. high conductance) synapses accelerates learning significantly.

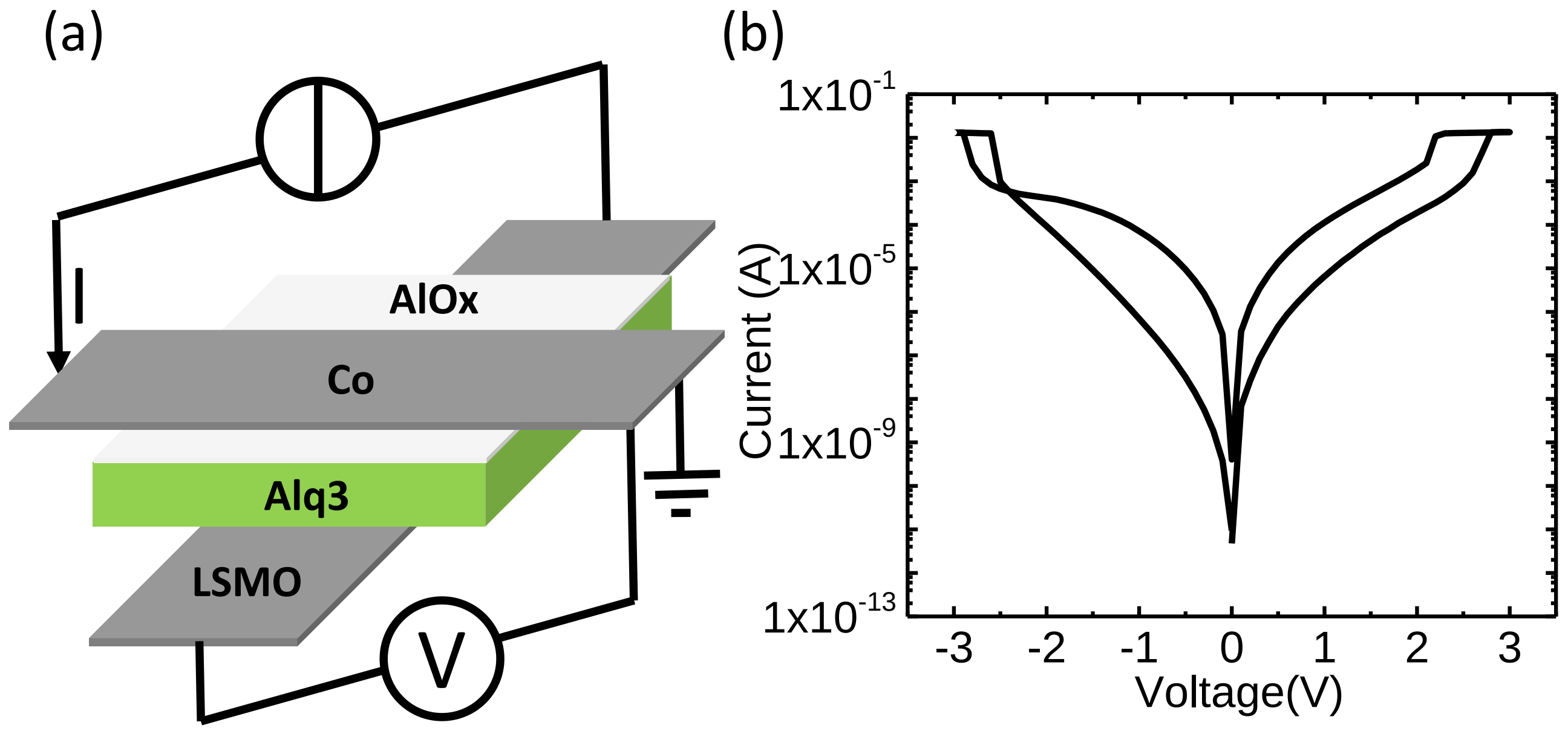

*Figure 1: (a) schematic diagram of the LSMO/Alq3/AlOx/Co molecular spin valves investigated in the present work. (b) Typical current voltage characteristic, that shows an hysteresis. This means that the device can act as a memory.*

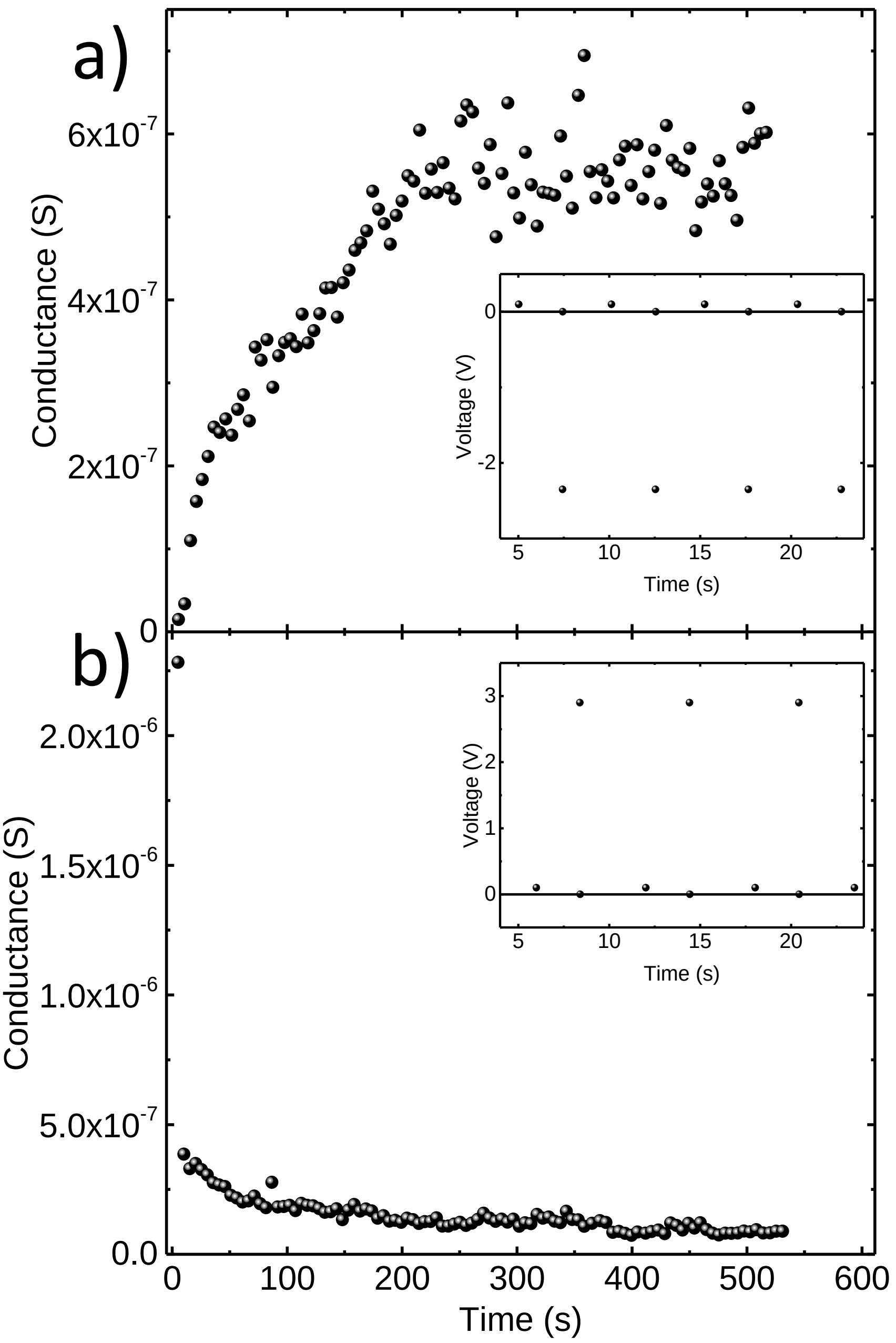

*Figure 2: Repeated, identical pulses can modify cumulatively the conductance of the device, which in neuromorphic computing plays the role of the synaptic weight. The conductance was measured at 100 mV, 100 K. a) After setting the device in a low conductance state by the application of -4 V voltage for 1 s, repeated pulses of 2.5 V lasting 5 ms were applied, causing the conductance to increase. b) After setting the device in a high conductance state by the application of 4 V voltage for 1 s, repeated -2.35 V pulses of 5 ms were applied, with a concomitant decrease of the conductance. The insets show the voltage sequence.*

The fundamental property of a synapse is that it can be potentiated (i.e. its weight is increased) or depressed (i.e. its weight is decreased) by the repeated application of a suitable signal. In the memristor embodiment of the synapse, the role of the weight is played by the conductance and that of the signal by a voltage pulse. Figure 1a) shows a schematic representation of the memristor devices studied in this work. The top Co electrode kept to ground; this means that the voltages in what follow are applied to the LSMO electrode relative to the Co one. Figure 1b) shows the hysteretic behaviour of the memristor's conductance. In particular, the application of a positive voltage increases the conductance (i.e. it potentiates the synapse), while the application of a negative one decreases it (i.e. it depresses the synapse). It must be borne in mind that only voltages that exceed $|V|>1.2$ can affect the conductance.

To demonstrate that the LSMO/Alq3/AlOx/Co device performed as a synapse, we applied repeated voltage pulses and measured its conductance at 0.1 V, which was small enough not to perturb its conductance. Figure 2a) shows a potentiation sequence, in which pulses of 2.5 V, lasting 5 ms were applied. Figure 2b) shows a depression sequence, obtained by applying -2.35 V pulses of 5 ms. This performance can be compared to that of NAND flash memories used for storage, which have write/erase times of the order of milliseconds, a $10^3$-$10^5$ erase cycle endurance and a 15 V write/erase voltage[19].

The ON/OFF ratio was calculated as the ratio between the final conductance and the initial one. The ON/OFF ratio of the potentiation sequence was 47 in this case, while that of the depression was 31. The conductance increase was rather gradual for the potentiation sequence (Figure 2a), while it was more abrupt for the depressing sequence (Figure 2b). The insets to Figure 2a) and Figure 2b) show the first few voltage pulses for the potentiation and depression sequences respectively.

In order to explore the parameter space of the potentiation and depression pulses, we applied pulse sequences of different duration and intensity. For the potentiation pulses, we first set the device in the low conductance state by applying a -4 V voltage for 1 s; subsequently 50 identical positive pulses were applied. This process was repeated for various pulse voltages and time durations and the results are reported in Figure 3a). The same algorithm was used to study the depression pulses, starting by setting the devices in the high conductance state by applying +4 V for 1 s; the results are shown in Figure 3b).

Previous results on similar devices only addressed the ability to set the conductance of a device to a desired value by the application of a voltage pulse of a specific amplitude[20], [21]. These experiments go a step further and demonstrate that the repeated application of identical voltage pulses can change the conductance of the memristor progressively.

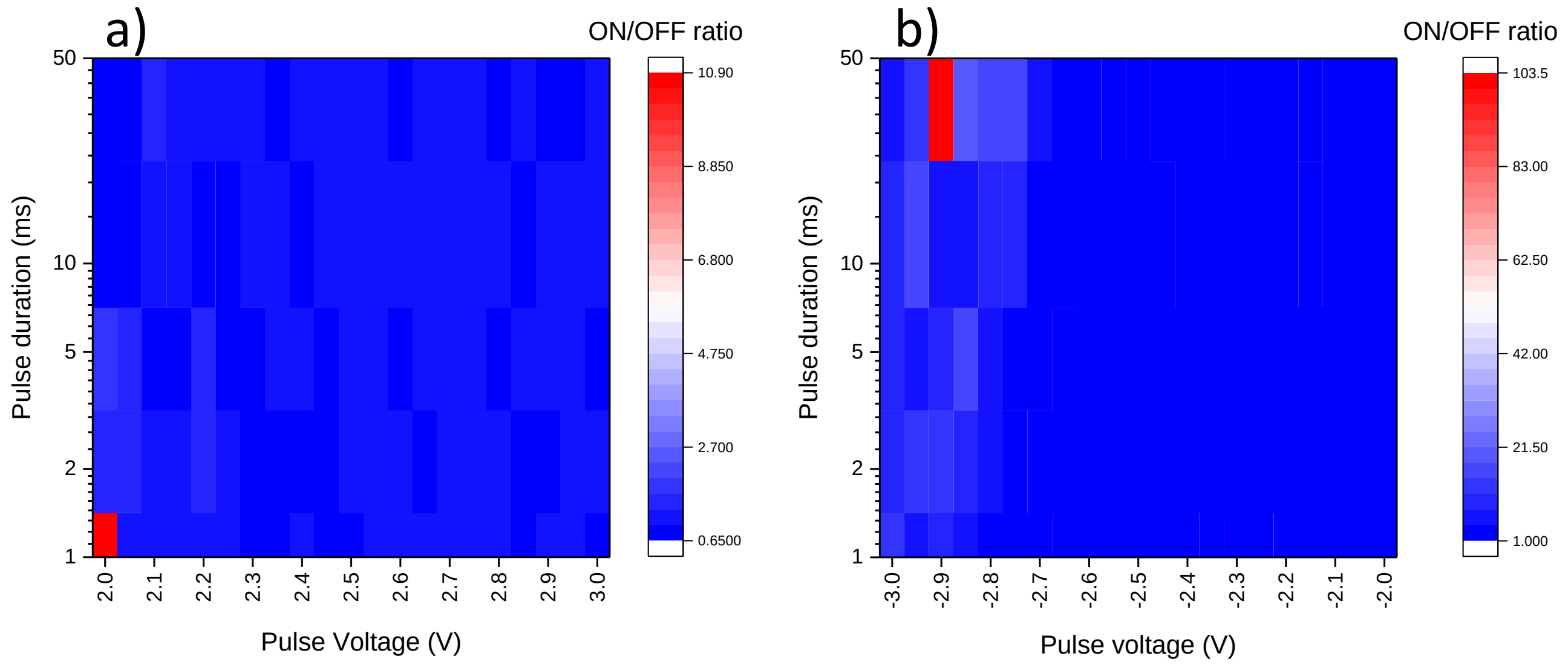

*Figure 3: On/Off ratio maps. a) The map reports the change in conductance from a low conductance sate, after the application of 50 potentiation pulses. The device was set in a low conductance state by the application of -4 V for 1s. Then 50 voltage pulses with duration and intensity shown in the axes were applied. b) Change in conductance starting from a high conductance state, after the application of 50 depression pulses. The device was set in a high conductance state by the application of 4 V for 1s. Then 50 voltage pulses with duration and intensity shown in the axes were applied.*

Compared to other solid state realizations of synapses, molecular spintronic devices have the additional ability to change their conductivity by controlling their ferromagnetic electrodes' magnetizations. This happens in a specific way: only the potentiated synapses can be further potentiated by setting the magnetizations of the electrodes in the antiparallel state, for which the conductance is maximum. The depressed synapses are not affected by the magnetic field and remain unpotentiated, i.e. their conductance does not change. This happens because the two phenomena are intertwined: the magnetic bistability is suppressed when the device is in the low conductance state and is maximum when the conductance is greatest.

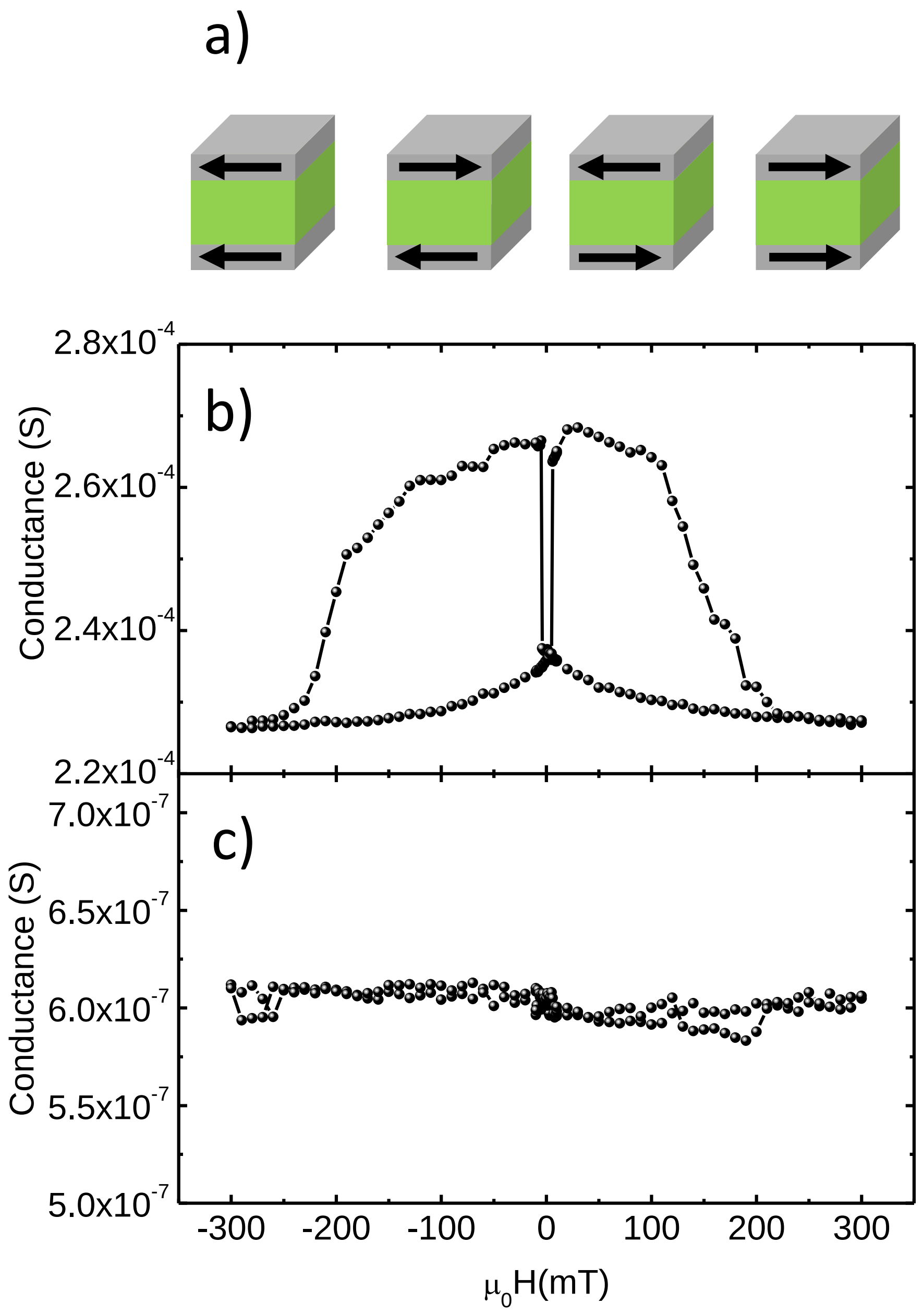

*Figure 4: In the high conductance state, the conductance of the device depends on the relative orientation of the magnetization of the ferromagnetic electrodes. a) The black arrows indicate the direction of the magnetization of the two electrodes. b) In a high conductance state, the conductance is highest when the magnetizations are oriented antiparallel to each other. c) In a low conductance state, the conductance does not depend on the magnetization of the electrodes.*

Since the magnetic bistability is due to the fact that the resistance of a device depends on the relative orientation of the magnetization of the electrodes, the magnetoconductance *MG* can be defined as:

$$MG = \frac{G_{AP} - G_P}{G_P}$$

were $G_{AP}$ is the conductance of the device when the magnetizations of the electrodes are antiparallel to each other and $G_P$ is the conductance for the parallel case. Figure 4a) shows the orientation of the magnetization of the top Co and bottom LSMO ferromagnetic electrodes. In the high conductance state, Figure 4b), the conductance was further increased by setting the electrodes in the antiparallel state. When the same magnetic field was applied to the device in the low conductance state, no change in its conductance was observed, as shown in Figure 4c). This behaviour illustrates how these devices can be expediently applied to neuromorphic computing: in one single step, high weight synapses can be globally potentiated, while leaving the low weight ones unaffected.

Besides these extremal behaviours, intermediate ones were achieved too as shown in Figure 5a), and could be modelled by a simple mathematical expression. The magnetoconductance as function of the conductance was 0 below a conductance threshold ($G_{th}$=1.13×10$^{-6}$ S in the case at hand) and followed a power law from this value up to the maximum conductance:

$$\begin{cases} MG(G) = 0 \; for \; G_{min} < G < G_{th} \\ MG(G) \propto (G - G_{th})^{\frac{3}{4}} \; for \; G_{th} \leq G < G_{max} \end{cases} \qquad (1)$$

For the device under study, $G_{min}$=6.4×10$^{-7}$ S and $G_{max}$=8.9×10$^{-5}$ S. For the application of such a device in a neuromorphic circuit, the synaptic weight was represented by the conductance $G$. If $G$ is the value of the conductance in the parallel magnetization configuration, and $G > G_{th}$, then by setting the device in the antiparallel magnetization state, the synaptic weight is boosted by an amount equal to $MG(G)$. It value was 25% at most in our case, while MG in excess of 604% were reported in literature for a different kind of spin valves[16].

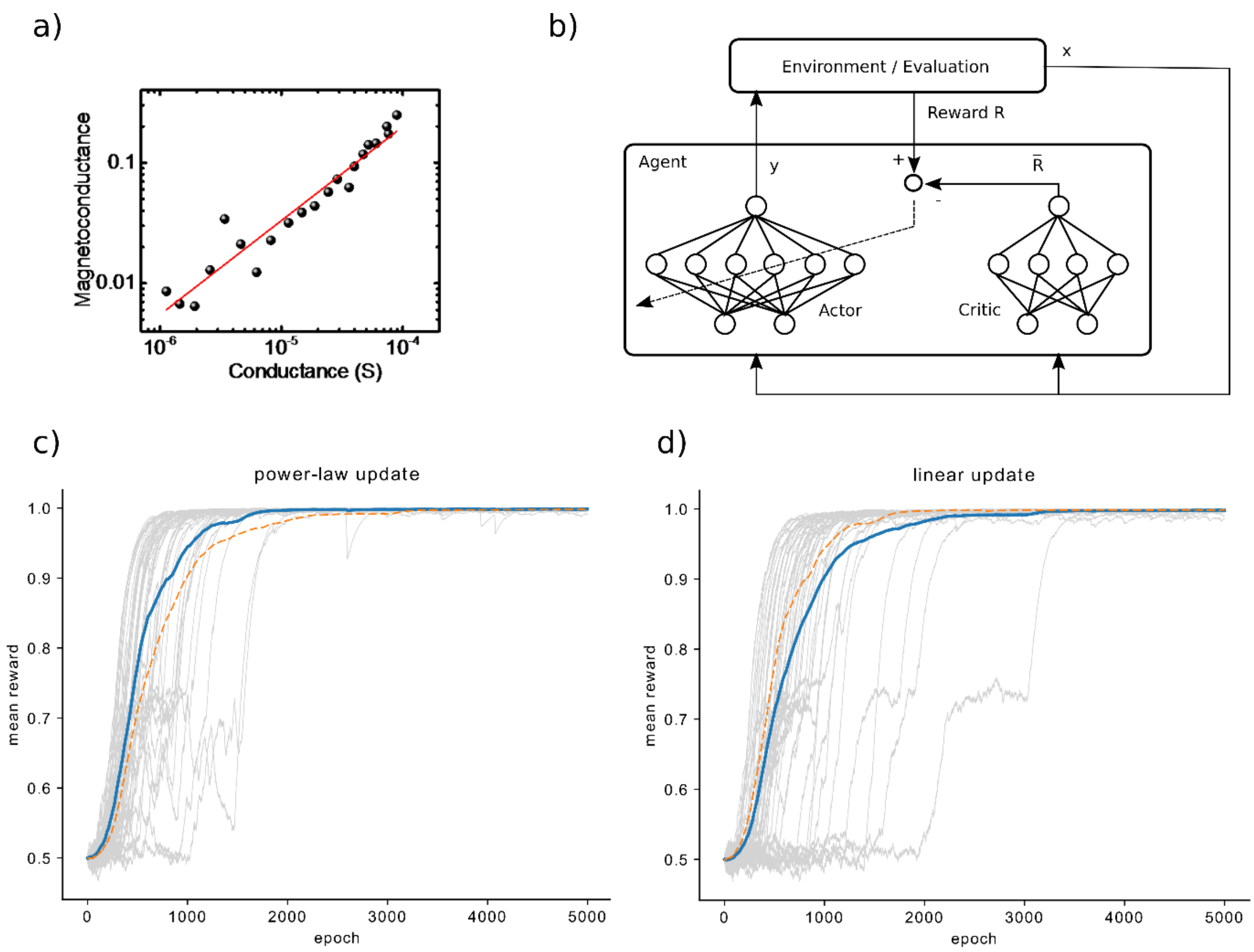

*Caption for Figure 5: Nonlinear parameter updates in an stochastic neural network. a) Dependence of the magnetoresistance on the conductance of a device. b) Actor-critic network schema for reward-based learning. The agent consists of two networks, and actor network (left) and a critic network (right). Upon an input x, the actor produces an output y which is evaluated and a reward R is provided by the environment. The critic network predicts the average reward for the given network input. The difference between the predicted reward and the actual reward is used to train the weights of the actor network. c,d) Average reward returned per epoch during training for a non-linear power-law update (c) and a linear update (d). Individual learning trials are shown in gray, the average performance over all trials is shown in blue. For comparison, the broken orange trace in (c) shows the trial-average for the linear update and vice versa in (d).*

The ability to globally boost the conductance of potentiated synapses in a nonlinear manner by the application of a magnetic field can speed up the leaning process in a neuromorphic circuit. Nonlinear updates have been proposed to be beneficial in reward-based learning as they can supress the effect of small noise-induced parameter updates while emphasizing stronger parameter changes in gradient directions that are consistent over many training examples[22], [23]. In order to test this for the nonlinear updates of the magnetoconductance, we considered reward-based learning in an actor-critic framework[24], see Figure 5b). In a reward-based learning setup, an agent receives an input $\boldsymbol{x}$ from the environment and reacts with an output $y(\boldsymbol{x})$. The output is evaluated and a reward $R$ is returned which measures the performance of the agent for this input. Then, another input is presented $\boldsymbol{x}$ (which may depend on the current state of the environment and the previous action of the agent), and so on. The actor-critic model considered here consists of two networks. An actor network learns a policy, that is, a suitable output $y(\boldsymbol{x})$ for the current input $\boldsymbol{x}$. Another network, called the critic network, learns the expected reward $\bar{R}(\boldsymbol{x})$ for this input and the output of the actor network. The critic network is used to improve learning of the actor network. We used for the actor network a stochastic binary neural network (see Methods for details) consisting of two input neurons, one hidden layer with 10 neurons, and one output neuron. The stochasticity of the network drives exploration, which is necessary in reward-based learning as the network has to explore which actions will lead to rewards. As the critic network, we used a standard sigmoidal neural network with one hidden layer consisting of 20 neurons. This network could in principle be trained with gradient descent. However, we simplified the weight update rules to obtain parameter updates that could be easily implemented in hardware (see Methods). The critic network predicts the mean reward $\bar{R}(\boldsymbol{x})$ achieved by the current actor network for the current input $\boldsymbol{x}$. This usually speeds up learning, but it is not necessary that the critic predicts the mean reward exactly, which justifies the use of simpler training procedures for this network.

Proposed weight updates (and updates for the biases) in the actor network are accumulated over batches (we used a batch size $10$ in the simulations) using a standard policy gradient update rule with a global reward signal[25]–[27]:

$$\Delta w_{ij}^{\mathrm{batch}} \leftarrow \Delta w_{ij}^{\mathrm{batch}} + \eta(R - \bar{R}(\boldsymbol{x})) \cdot \big(y_i(\boldsymbol{x}) - p_i(\boldsymbol{x})\big) y_j(\boldsymbol{x}), \tag{2}$$

where $\eta > 0$ is a small learning rate $\Delta w_{ij}^{\mathrm{batch}}$ is the change of the weight from neuron $j$ to neuron $i$, which is accumulated over the current batch. $y_i(\boldsymbol{x})$ is the output of neuron $i$ for input $\boldsymbol{x}$, $p_i(\boldsymbol{x})$ is the probability that the neuron gives this output, and $\bar{R}(\boldsymbol{x})$ is the expected reward provided by the critic network. After a batch, these accumulated updates are added to the weights as described below and $\Delta w_{ij}^{\mathrm{batch}}$ is reset to 0. The nonlinear update of the weights is given by a power law according to the memristive conductance change due to a magnetic pulse:

$$\Delta w_{ij} = \begin{cases} \mathrm{sign}\left(\Delta w_{ij}^{\mathrm{batch}}\right)\left|\Delta w_{ij}^{\mathrm{batch}}\right|^{\frac{7}{4}} & \text{if } \left|\Delta w_{ij}^{\mathrm{batch}}\right| > \Delta w^{\mathrm{min}} \\ 0 & \text{otherwise} \end{cases}. \quad (3)$$

Hence, weight changes are only implemented if the accumulated change is above a threshold $\Delta w^{\mathrm{min}}$ (we used an update threshold of 0.4 in the simulations below). In this case, the accumulated change is transformed by a power-law that preserves the sign of the change. We compared this nonlinear update with a linear update where weights were we used $\Delta w_{ij} = \Delta w_{ij}^{\mathrm{batch}}$ after each batch.

We tested this model on a logical XOR task. The XOR task is a simple but standard task for reward-based learning with a global reward signal. Since it is nonlinear, it cannot be solved by a single neuron, and reward based learning with a global signal tends to get stuck in local optima. The input to the network consisted of two binary values $\boldsymbol{x} \in \{0,1\}^2$ and the target output is XOR of the two bits. A returned reward was 1 if the actor network output matched the target and 0 otherwise.  The average reward over learning epochs (an epoch is defined as one batch) for the nonlinear update are shown in Figure 5c) and compared to a linear update in Figure 5d). To obtain the learning curves, we performed a parameter search over learning rates. Figure 5c and 5d show the results with the best-performing learning rate for the nonlinear and linear updates respectively. The nonlinear update performed significantly better (50 learning trials; Epochs to reach a performance goal of 0.975 was 896±301 in the nonlinear case and 1076±484 in the nonlinear case; Welch's t-test: $p<0.05$). More interestingly, the linear update resulted in more learning trails with very long convergence times, which was largely absent in the nonlinear update (gray traces in Figure 5c, d).

**Conclusions**

We have demonstrated that LSMO/Alq3/AlOx/Co molecular spin valves can be used as synaptic weights in neuromorphic circuits. We were able to progressively increase or decrease their conductance by the repeated application of voltage pulses, which is necessary for neuromorphic computing. We also demonstrated that their conductance could be controlled by the relative orientation of the magnetization of the two ferromagnetic electrodes. This magnetoconductance depended on the conductance of the device and we modelled its behaviour with a simple mathematical law. We used this equation as a nonlinear update rule in the simulation of an actor/critic neuromorphic circuit. We chose the learning of the XOR function as a test problem, which led to better convergence properties and faster learning.

**Acknowledgments**: A.R. would like to acknowledge fruitful discussions with Dr.Mirko Prezioso.

**Funding Acknowledgements:** This work was supported by the European Union Seventh Framework Programme (FP7/2007-2013) project Organic-inorganic hybrids for electronics and photonics (HINTS), under grant agreement GA no. 263104, and by the Austrian Science Fund (FWF): I 3251-N33.

**Methods**

Actor network:

We used for actor network consisted of two input neurons, fully connected to a single hidden layer consisting of 20 stochastic binary neurons. Finally, the hidden layer was fully connected to one stochastic binary output neuron. To simplify the description of stochastic binary neurons, we consider a neuron that receives input $\boldsymbol{x}$. Each neuron $i$ computes a probability of its output being one: $p_i = \sigma(\boldsymbol{w}_i^T \cdot \boldsymbol{x} + b_i)$ where $\sigma$ is the logistic sigmoid function, $\boldsymbol{w}_i$ is the weight vector and $b_i$ the bias for neuron $i$. A random bit $y_i^{proposed}$ is drawn from a Bernoulli distribution with parameter $p_i$ (i.e., $y_i^{proposed} = 1$ with probability $p_i$). This is not yet the output of the neuron. In order to improve exploration at states where the expected reward is low, the output bit can be flipped. In particular, each neuron computes a flip probability $p_i^{flip}(\boldsymbol{x}) = \alpha^{flip}(1 - \bar{R}(\boldsymbol{x}))$, for some $\alpha^{flip} \in [0,1]$ and $\bar{R}(\boldsymbol{x})$ being the expected reward provided by the critic network, where we assume that the expected reward is smaller or equal to 1. The output is then given by $y_i(\boldsymbol{x}) = 1 - y_i^{proposed}$ with probability $p_i^{flip}(x)$ and $y_i(\boldsymbol{x}) = y_i^{proposed}$ with probability $1 - p_i^{flip}(x)$.

Network parameters: The network had two input neurons, 10 hidden neurons, and one output neuron. All layers were fully connected. The bit-flip factor was set to $\alpha^{flip} = 0.1$. Weights were initialized uniformly over $\left[-\frac{1}{\sqrt{N}}, \frac{1}{\sqrt{N}}\right]$, where $N$ is the number of inputs to the neuron. Biases were initialized to 0. Training was done with a batch size of 10 with an update threshold $\Delta w^{\min} = 0.4$. A parameter search over learning rates was performed. Learning rates for hidden layer neurons were varied from 0.4 to 1.25 in steps of 0.05. The learning rate of the output neuron was set to half the learning rate of the hidden neurons. The best hidden-layer learning rate was 0.75 for the linear update and 1.1 for the nonlinear update.

Critic network:

The critic network was a standard sigmoidal neural network with one 10-neuron hidden layer. The activation functions of the hidden neurons and the output neuron were logistic sigmoids. The weights to the output neuron were fixed and not changed during training. Weights to the hidden neurons were updated through gradient descent to minimize the sum-squared error of the network output with the current reward as the target. Note that with output weights fixed, this results in a rule that needs the difference between the network output and the current reward as the only global learning signal.

Network parameters: The network had two input neurons, 20 hidden neurons, and one output neuron. All layers were fully connected. Hidden layer weights were initialized uniformly over $\left[-\frac{1}{\sqrt{N}}, \frac{1}{\sqrt{N}}\right]$, where $N$ is the number of inputs to the neuron. Biases were initialized to 0. Output neuron weights were initialized uniformly over [-1.25, 1.25] with biases initialized to 0.5. The network was trained with a batch size of 1 and a learning rate of 1. Only the hidden layer was trained, the output neuron weights were not updated during training. The weight update for a hidden layer weight was given by

$$\Delta w_{ij} = (R - y^{out}) y_i w_i^{out} y_j - 0.001\mathrm{sign}(w_{ij}),$$

Where $y^{out}$ denotes the network output, $y_i$ is the output of neuron $i$ and $w_i^{out}$ is the weight from hidden neuron $i$ to the output neuron. Last term implements an L1-norm regularization of hidden layer weights.

The reward-traces shown in Figure 5c,d were filtered using the online-exponential filter $R_{\text{filtered}}(n) = 0.999\, R_{\text{filtered}}(n-1) + 0.001\, R(n)$, where $R(n)$ is the reward returned after the n$^{\text{th}}$ presentation of a training example.